\documentclass[review]{elsarticle}
\usepackage[utf8]{inputenc}
\usepackage[T1]{fontenc}
\usepackage{amsmath}
\usepackage{url}

\usepackage{lineno}
\modulolinenumbers[5]

\journal{Journal of Magnetic Resonance Open}
\usepackage{float}

\begin{document}

\begin{frontmatter}
\renewcommand\thesubsection{\Alph{subsection}}
\renewcommand\thesubsubsection{\thesubsection.\Roman{subsection}}

\title{Geometry dependence of micron-scale NMR signals on NV-diamond chips }

\author{F. Bruckmaier\fnref{TUM}}
\author{K. Briegel\fnref{TUM}}
\author{D. B. Bucher\fnref{TUM}\corref{correspondingAuthor}}
\address{Department of Chemistry, Technical University of Munich\\Lichtenbergst 4, 85748 Garching, Germany}

\cortext[correspondingAuthor]{Corresponding author}
\ead{Dominik.Bucher@tum.de}
\fntext[TUM]{Department of Chemistry, Technical University of Munich, Germany}

\begin{abstract}
Small volume nuclear magnetic resonance spectroscopy (NMR) has recently made considerable progress due to rapid developments in the field of quantum sensing using nitrogen vacancy (NV) centers. These optically active defects in the diamond lattice have been used to probe unprecedented small volumes on the picoliter range with high spectral resolution. However, the NMR signal size depends strongly on both the diamond sensor's and sample's geometry. Using Monte-Carlo integration of sample spin dipole moments, the magnetic field projection along the orientation of the NV center for different geometries has been analyzed. We show that the NMR signal strongly depends on the NV-center orientation with respect to the diamond surface. While the signal of currently used planar diamond sensors converges as a function of the sample volume, more optimal geometries lead to a logarithmically diverging signal. Finally, we simulate the expected signal for spherical, cylindrical and nearly-2D sample volumes, covering relevant geometries for interesting applications in NV-NMR such as single-cell biology or NV-based hyperpolarization. The results provide a guideline for NV-NMR spectroscopy of microscopic objects.

\end{abstract}

\begin{keyword}
 Nitrogen vacancy center\sep nuclear magnetic resonance\sep Monte-Carlo\sep quantum sensing\sep sample geometry \sep small volume NMR
\end{keyword}

\end{frontmatter}

\flushbottom

\thispagestyle{empty}

\section*{Introduction}

Nuclear magnetic resonance spectroscopy (NMR) is one of the most powerful analytical tools for molecular structural analysis, however, it is limited by its intrinsically low sensitivity, and often restricted to large sample mass (mg) or volume ($\mu$l). Microcoils have significantly reduced the sample size of conventional NMR experiments down to $0.1$ nL volumes \cite{Fratila2011,Badilita2012,Lepucki2020,Grisi2017}. The main challenges to further reduce the sample volume lie in the difficulty of fabricating the microscopic pick-up coils and the high sensitivity required to detect a signal from an even smaller number of spins \cite{Fratila2011}.
These problems can be overcome with a novel diamond based quantum sensor: the nitrogen vacancy (NV) center \cite{Schirhagl2014}.
This atomic sized sensor can be brought in close proximity to the sample and has demonstrated superior spin number sensitivity ranging from detection of molecular sub-monolayers at surfaces \cite{liu2021surface} down to a single molecule \cite{Lovchinsky836} or even a single nuclear spin \cite{Mller2014,Sushkov2014}.
The actual detection volume of any given NV-NMR diamond chip strongly depends on the depth of NV-centers below the diamond surface which divides NV-NMR in two distinct subgroups: i) Nanoscale NV-NMR based on single or an ensemble of NV centers located a couple of nanometers below the diamond surface \cite{Mamin2013,Staudacher561}. Due to the small number of sample spins in the nanoscale detection volume, the NV center detects magnetic polarization caused by random fluctuations within the spin bath (statistical polarization) \cite{meriles_API_2010}. ii) Micron-scale NV-NMR based on a few micrometer thick NV-layer at the diamond surface is able to detect microscopic sample volumes, where the thermal polarization typically exceeds the statistical polarization \cite{Glenn2018,Arunkumar2021,Smitseaaw7895,Bucher2020,Kehayias2017}. These sensors are of particular interest since their sensing volume is on the order of picoliters, which corresponds to the volume of a single mammalian cell. Ultimately this may allow for non-invasive and non-destructive detection of metabolites in single cells \cite{Fessenden2016}.  \\
In contrast to induction based NMR techniques, where the principle of reciprocity is typically applied to estimate the signal strength based on the detection coil geometry \cite{Hoult1976}, in the NV case this can be achieved by integrating over the spin magnetic moments in the sample geometry. The origin of the NMR signal and the signal's dependence on sample geometry has been investigated for shallow NV centers, exposing a strong dependence on the NV-depth and a weaker angular dependence \cite{Mamin2013,Staudacher561,meriles_API_2010,Phan_physRevB_2016,Merilesgeometry}. 
Here, we show that for micron-scale NV-NMR the geometry of the diamond and sample is pivotal. We first derive an analytical model of a sensitivity map for thermally polarized signals following Glenn et al. \cite{Glenn2018}. Then we compare these with simulations performed via numerical Monte-Carlo integration and suggest different ways to optimize the diamond geometry. Finally, we examine the signal of various sample geometries of particular interest for future research.

We would like to point out that similar models have been used in NMR spectroscopy for a in depth analysis of flux densities from the macroscopic sample magnetization at an individual nuclear site, which can cause artefacts in high density nuclear spin baths \cite{LevittMacroMicro2,VlassenbroekMakroMicro}. The dipolar coupling of a nuclear spin bath has also been used to detect NMR signals \cite{VathyamSQNMR,BowtellDipolarFieldImaging,GranwehrXenonNMR,DongOilWater}, which is related to the NV-based detection scheme. However, most of these results are not applicable to micron-scale NV-NMR spectroscopy due to the different geometry requirements of the NV-center and the experimental setup. \\


\section*{Basic geometry of an NV-NMR experiment}
\begin{figure}[H]
\centering
\includegraphics[width=\linewidth]{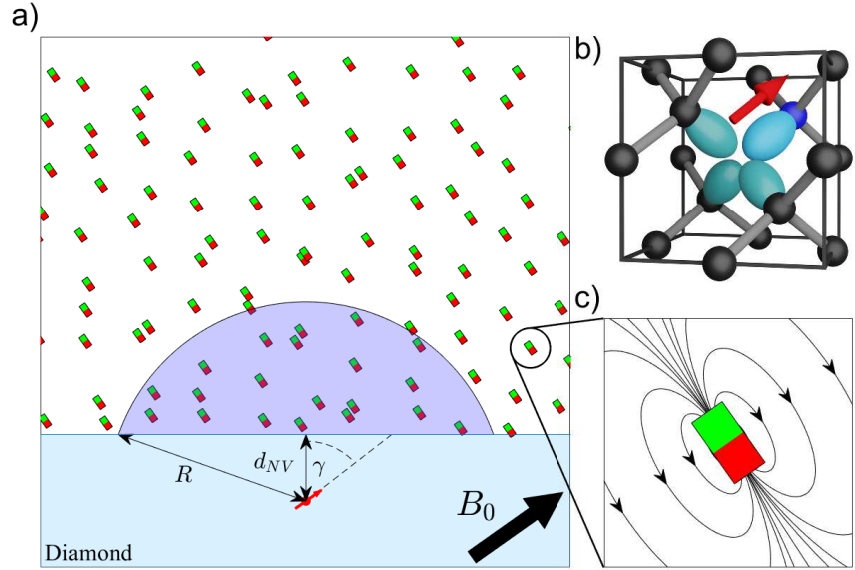}
\caption{\textbf{a)} Schematic of a typical NV-NMR experiment. The diamond is depicted as the pale blue area with a red arrow indicating the NV center and its orientation at a depth $d_{NV}$ below the diamond surface. The blue circular segment of radius $R$ is the area out of which $> 50\%$ of the NMR signal originates. The miniature bar magnets depict the polarized nuclear sample spins after a $\pi/2$ pulse and the black arrow depicts the external magnetic field $B_0$ parallel to the NV center orientation. \textbf{b)} The diamond crystal lattice with nitrogen, depicted in blue and an adjacent vacancy in the center of the cube, forming an NV center. \textbf{c)} Miniature bar magnet depicting a magnetic dipole and a schematic of its field lines.}
\label{fig:Setup}
\end{figure}

NV centers in diamond used for sensing thermally polarized spins in microscopic sample volumes are typically located in a layer of a few $\mu$m thickness at the diamond surface \cite{Glenn2018,Arunkumar2021,Smitseaaw7895,bucher_quantum_2019,Kehayias2017}. As in conventional NMR experiments, the measured signal originates from the polarized spins in the sample, which will precess at their Larmor frequency $\omega_L$ around an external magnetic field $B_0$ after applying a $\pi/2$ pulse.\\
The geometry of an NV-NMR experiment is depicted in Figure \ref{fig:Setup}a) for a single NV center. Due to the tetrahedral nature of the diamond crystal lattice, shown in Figure \ref{fig:Setup}b), there will be four different orientations of NV centers in the crystal, one along each of the <111> axes. Typically, the  $B_0$ field is aligned along one NV-orientation \cite{bucher_quantum_2019}, which is then used for sensing applications. Since most diamond surfaces are cut along the <100> axis, this leads to a $\sim$54.74$^\circ$ angle of both the magnetic field and the NV center orientation to the surface normal. For simplicity, we discuss the geometry for the single NV case in the main paper. The more important NV-ensemble case behaves qualitatively similar and can be found in the supplementary material section \ref{sec:NVlayers}.

\section*{Analytical derivation of the NMR signal}
The magnetic field of a single precessing nuclear spin at point $\vec{r}$, such as from a proton in our sample, is depicted in Figure \ref{fig:Setup}c) and can be mathematically expressed as:

\begin{equation}\label{eq:start}
    \vec{B}_{\mathrm{spin}}(\vec{r},t) = \frac{\mu_0}{4\pi} \left[ \frac{3 \hat{r} \left( \hat{r} \cdot \vec{m}(t) \right) -  \vec{m}(t)}{r^3} \right]
\end{equation}

Where $\mu_0$ is the magnetic permeability, $\hat{r}$ is the direction and $r = |\vec{r}\,|$ the distance from the sample spin to the NV center and $\vec{m}(t)$ is the magnetic moment of the precessing nuclear spin. Since in typical NMR sensing pulse sequences the NV center is only sensitive to magnetic fields along $\hat{B}_0$ \cite{meriles_API_2010,bucher_quantum_2019,Degen2017}, we are interested in the projection of $\vec{B}_{\mathrm{spin}}$ onto the NV and magnetic field orientation $\hat{B}_0$. Assuming a uniform polarization of the sample, we can integrate over the volume of a spherical cap of radius R above the diamond surface resulting in the total magnetic field strength S sensed by the NV center:

\begin{equation}\label{eq:signal}
\begin{split}
    S &= \int \rho_{\mathrm{therm}} \,  \left(\vec{B}_{\mathrm{spin}}(\vec{r},t) \cdot \hat{B}_{0}\right) \, dV \\
    &= \frac{\mu_0}{4\pi} \, \rho_{\mathrm{therm}} \,\overline{|m(t)|} \int  \,  \left[ \frac{3 \hat{r} \left( \hat{r} \cdot \hat{m}_{\mathrm{max}} \right) -  \hat{m}_{\mathrm{max}}}{r^3}\right] \cdot \hat{B}_{0} \, dV 
\end{split}
\end{equation}

Where $\rho_{\mathrm{therm}} $ is the concentration of thermally polarized sample spins, $ \hat{m}_{\mathrm{max}}$ is a spin orientation where $\hat{m}$ lies in the plane defined by $\hat{B}_0$ and the surface normal, this being the orientation of maximum signal amplitude and $\overline{|m(t)|}$ is the mean amplitude of the spin magnetic field projection along $\hat{B}_0$ at the NV location. From this, we can see that the signal S can be split up into two parts $S = K \cdot G$, where $K = \frac{\mu_0}{4\pi} \, \rho_{\mathrm{therm}} \,\overline{|m(t)|} $ is a geometry independent prefactor and gives an approximate scale of the magnetic field strength, whereas the geometry factor $G =\int  \,  \left[ \frac{3 \hat{r} \left( \hat{r} \cdot \hat{m}_{\mathrm{max}} \right) -  \hat{m}_{\mathrm{max}}}{r^3}\right] \, dV $ is purely dependent on the sample's geometry. Integrating G with respect to the geometry depicted in Figure \ref{fig:Setup}a), results in:

\begin{equation}\label{eq:G factor}
   G =  \pi \,\sin{\gamma}\,\cos{\gamma} \, \frac{2 R^3 -3 d_{NV} R^2 +d_{NV}^3}{R^3}
\end{equation}

Where $\gamma$ is the angle between the NV orientation and the surface normal and $d_{NV}$ denotes the depth of the NV center. A more detailed derivation of Equation \ref{eq:G factor} is provided in the supplementary material section \ref{sec:geometryfac}. Taking the limit of this equation for an infinite sample volume results in:

\begin{equation}\label{eq:G infty}
   G_{\infty} =  \lim_{R\to\infty}  \pi \,\sin{\gamma}\,\cos{\gamma}\,\frac{2 R^3 -3 d_{NV} R^2 +d_{NV}^3}{R^3}  = 2 \pi\, \sin{\gamma}\,\cos{\gamma}
\end{equation}

\begin{figure}[H]
\centering
\includegraphics[width=1\textwidth]{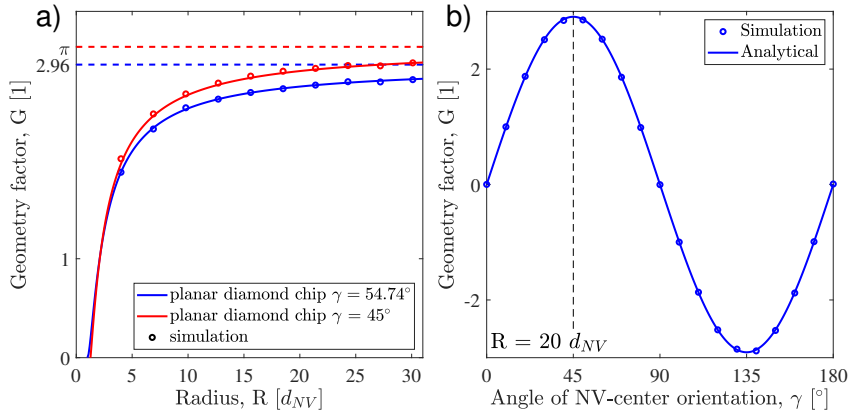}
\caption{\textbf{a)} Analytic plots (lines) and numerical simulation (circles) of the geometry factor seen by the NV center for a planar diamond chip versus the radius of the sample volume with the angle between the surface normal and $\hat{B}_0$ at $\gamma = 54.74^\circ$ and $\gamma = 45^\circ$. The geometry factor for both planar geometries converges with increasing radius R. \textbf{b)} The geometry factor of a planar diamond chip as a function of the angle between the surface normal and $\hat{B}_0$. The function follows the form $\sin{\gamma}\cos{\gamma}$ and has its maximum at 45$^\circ$. The radius for the spherical cap in this case is $R=20 \, d_{NV}$.}
\label{fig:Gfactor}
\end{figure}

 This has three important consequences: 
 i) The NMR signal converges as a function of R, which sets the detection volume of the NV-center (depicted in Figure \ref{fig:Gfactor} a). 
 ii) For large sample radii, the size of the signal is independent of the NV center's depth.
 iii) The signal size is dependent on the orientation of the NV center with respect to the surface captured in $\gamma$ (i.e. the diamond cut). The maximum NMR signal at the NV position is reached for a 45$^\circ$ angle, with a geometry factor of $G = \pi$, whereas for an NV center orientated parallel to the surface the signal will be zero (Figure \ref{fig:Gfactor} b). Thus, the diamond cut must be considered carefully when thermal polarisation is to be detected. Note, that ii) and iii) are very different from the nano-scale NV-NMR, where the signal has a weak angular and strong dependence on $d_{NV}$. These analytical results are reproduced by Monte-Carlo simulations, also depicted in Figure \ref{fig:Gfactor} which are discussed in more detail in the supplementary material section \ref{sec:MonteCarlo}.
 
\begin{figure}[H]
\centering
\includegraphics[width=\linewidth]{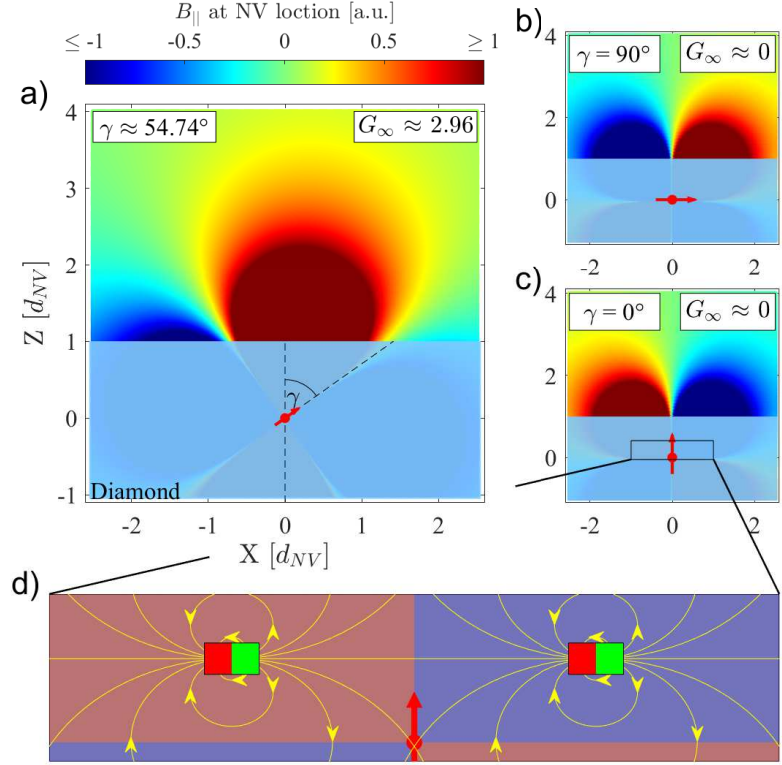}
\caption{Visualization of the origin of the NMR signal sensed by the NV (red arrow). For each point the magnetic field contribution at the NV position is projected along $B_0$. Red areas describe positive signals and blue areas negative signals at the NV-position. 
\textbf{a)} $\hat{B}_0$ is at an angle of 54.74$^\circ$ relative to the surface normal, as in most NV-NMR experiments, resulting in $G_{\infty} = 2.96$.
\textbf{b)-c)} $\hat{B}_0$ is at an angle of 90$^\circ$ and 0$^\circ$ respectively, resulting in a cancellation of the positive and negative signal due to symmetry and therefore a total geometry factor of 0.
\textbf{d)} The dipoles in the blue and red area will have the same magnitude but opposite direction of the magnetic field when projected on $\hat{B}_0$ at the NV location, therefore the total geometry factor in b) and c) is zero.}
\label{fig:quasiOrbital}
\end{figure}

These findings can be explained by analyzing the spatial contribution of each sample spin to the total signal at the NV position. This can be visualized by colour-coding space according to the sign and strength of the contribution to the geometry factor as shown in Figure \ref{fig:quasiOrbital}a). The space around the NV can be divided into four quadrants, with negative and positive signal contributions at the NV position. This leads to a total cancellation of the thermally polarized NMR signal for an NV-center immersed in a sample, e.g. a nano-diamond in water \cite{Holzgrafe2020}. However, for a diamond chip, some of this volume is occupied by the diamond itself, breaking the symmetry and leaving a nonzero signal - which is the strongest for $\gamma = 45^\circ$. It is easy to see, why diamond chips cut along <111> (<110>) with NV centers parallel (orthogonal) to the surface normal cannot be used, even though the experimental setup would be much easier: the geometry factor is zero since the magnetic field lines are cancelled at the NV position which is depicted in Figure \ref{fig:quasiOrbital} d).

 Assuming an angle of $\gamma = 54.74^\circ$, this limits the maximum signal achievable for the previously discussed setup to $S = K\,2\sqrt{2} \pi / 3$. For a water sample at temperature $T = 300$ K, $B_0 = 0.2 $ T and $\overline{|m(t)|} = \hbar\gamma_p/\pi$ the geometry independent term K can be calculated to be:
 \begin{equation}
     K = \frac{\mu_0}{4\pi}\;\frac{\hbar\gamma_p}{\pi}\;\frac{\gamma_p \hbar \mathrm{B}_0}{2 \,k_B T}\;\rho_{\mathrm{proton}}\approx 80 \, pT
 \end{equation}
 Where $\gamma_p$ is the proton gyromagnetic ratio, $\hbar$ is the reduced Planck constant, $k_\mathrm{B}$ is the Boltzmann constant and $\rho_{\mathrm{proton}}$ the concentration of protons in water. Combined with G$_\infty$, this results in an estimated maximum signal of $S\approx240\,\mathrm{pT}$.

\section*{Sample geometry factor estimation for different applications}
\begin{figure}[]
\centering
\includegraphics[width=\linewidth]{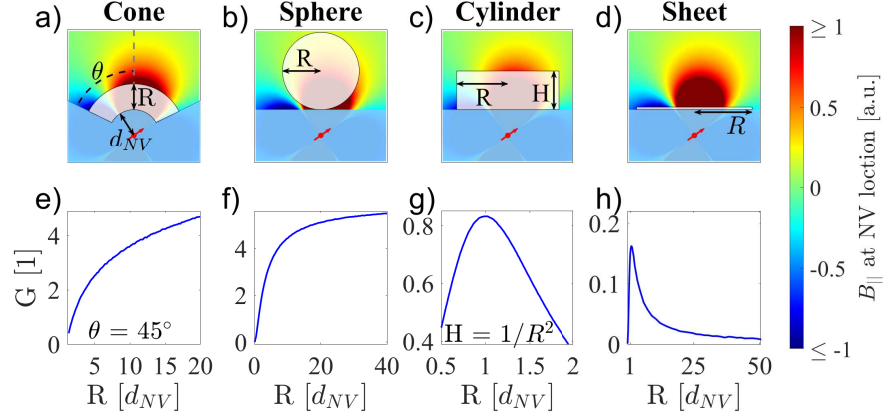}
\caption{\textbf{Cone: a)}  Schematic of cone shaped sample volume, the most noticeable improvement over the planar chip. It depends on the depth of the NV center, $d_{NV}$, the radius of the sample volume R and the angle of the cone $\theta$. \textbf{e)} Numerical simulation of the geometry factor of a cone-shaped sample for $\theta = 45^\circ$ and $d_{NV} = 1$. In contrast to the planar diamond, the geometry factor diverges logarithmically.\\
\textbf{Sphere: b)} A spherical sample on a planar NV diamond. By placing the sphere above the NV, the areas of negative signal can be avoided and therefore the overall signal increases. \textbf{f)} Numerical simulation of the change in the geometry factor with increasing radius.\\
\textbf{Cylinder: c)} Schematic of a cylindrical sample on a planar diamond chip. \textbf{g)} The geometry factor of the cylindrical volume. The total volume of the cylinder is kept constant and the height of the cylinder goes as $H \sim 1/R^2$ to simulate a cell's (near) constant volume. In this case, the maximum geometry factor is reached at $H=\sqrt{R}$.\\
\textbf{Nearly-2D sheet: d)} Schematic of a 0.1 $d_{NV}$ thin, uniformly polarized sheet on top of a planar diamond chip. It is defined through the distance to the NV center as well as the sheets radius. \textbf{h)} Numerical simulation of the geometry factor as a function of the sheet's radius.}
\label{fig:geometries}
\end{figure}

Published micron-scale NV-NMR experiments \cite{Glenn2018,Arunkumar2021,Smitseaaw7895,Bucher2020,Kehayias2017} have all used the geometry of the experiment as described in Figure \ref{fig:Setup}a), with a planar chip and an (approximately infinite) half-space filled with a sample. For further developments two major questions need to be answered. i) Which geometries maximize the geometry factor for optimal sensitivity? ii) What is the geometry factor for microscopic objects? This section will investigate some geometries of particular interest and their influence on signal strength compared to the standard geometry (Figure \ref{fig:Setup}a)). The various geometries (cone, sphere, cylinder and nearly-2D sheet) discussed here are depicted and summarized in Figure \ref{fig:geometries}a)-d). They all can be qualitatively understood by considering the spatial dependence of the NMR signal shown in Figure \ref{fig:quasiOrbital}. More detailed information on the simulations can be found in the supplementary material section \ref{sec:MonteCarlo}.

\subsection*{Cone}
A consequence of the quadrants' shape in Figure \ref{fig:quasiOrbital} a) is, that removing the volume with negative signal contribution increases the geometry factor. This results in a cone-shaped sample volume defined by the cone's angle $\theta$ and the radius of detection R, as described in Figure \ref{fig:geometries}. It is pretty straightforward to calculate this geometry factor analytically, as shown in the supplementary material section \ref{sec:geometryfac}. In contrast to the filled hemisphere considered before, $G_{\infty} $ does not converge anymore, but rather grows logarithmically with radius R. This signal can even be increased by removing the rotational symmetry about the z-axis, and filling in the whole quadrant above the diamond surface, in which the geometry-factor is positive. 

\subsection*{Sphere}
A fundamental microscopic sample geometry is the sphere. Samples of interest to which such a model would apply are non-adherent cells such as mammalian suspension cells, which are typically of $\sim$10 $\mu$m diameter \cite{gillooly2015nuclear}.
Another important system is microdroplets in microfluidics - water droplets surrounded by fluorinated oil. Due to their small volume, robustness in size and controllability on a chip, microdroplets are widely utilised for various applications in life sciences \cite{Kelly2007}. Spherical shapes are also able to increase the geometry factor by avoiding the areas of negative field as seen in Figure \ref{fig:geometries} b), though the total geometry factor will converge for increasing R. Our simulations indicate, that optimizing the size of the sphere as a function of NV-center depth can increase the geometry factor to $\sim$ 1.6 times that of the planar diamond chip geometry.

\subsection*{Cylinder}
While suspension cells are typically spherical, adherent cells can be crudely approximated by a cylindrical shape \cite{cylinderApprox1995,cylinderApprox2015}. A simulation of a cylinder with varying radius and fixed volume can be seen in Figure \ref{fig:geometries} c). For flat cells, the geometry factor can become very small. However, using this approximation will result in a lower bound for the geometry factor of adherent cells, since there will be more sample volume close to the cell's nucleus, which likely coincides with the region where the NV is most sensitive in, since one usually has control over the laser's position on the diamond. This is not taken into account in these basic geometry simulations, since the nucleus itself is of different size for different cells. As seen in Figure \ref{fig:geometries_layer}, this geometry benefits the most from the use of an ensemble of NV-centers.

\subsection*{Nearly-2D sheet}
Another interesting application is micron-scale NV-based NMR detection and hyperpolarization \cite{Rizzato2021,hyper_Healey,tetienne_prospects_2021}. One possible implementation is based on a diamond chip with a few micrometer thick NV layer, where the near-surface NV-centers are used to transfer polarization to sample nuclei at the diamond-sample interface and the full NV-ensemble detects the NMR signal. In the case of a liquid sample, these hyperpolarized sample spins will typically spread out over tens of $\mu$m due to molecular diffusion ($\sim$1e-9 $m^2$/s). For a diamond with $d_{NV}$ much smaller then the diffusion length the results discussed in Figure \ref{fig:Gfactor} and \ref{fig:quasiOrbital} are still valid. However, for a solid sample, spin diffusion will lead to a comparatively thin ($\leq \,1\, \mu$m) \cite{Merilesgeometry,Rizzato2021,hyper_Healey,BroadwaySurfaceHyperpol} sheet of  of hyperpolarized spins, in some cases smaller than $d_{NV}$. In our strongly simplified model, the sheet is approximated as a thin cylinder of varying radius. From the simulations we obtain a comparatively low geometry factor (Figure \ref{fig:geometries} h)), which reaches its maximum value at about $\sim$10 $\%$ of $G_\infty$. Our results show that the diamond cut and the overall geometry must be carefully chosen for NV-based hyperpolarization and detection experiments \cite{Merilesgeometry,BroadwaySurfaceHyperpol}.
However, for a quantitative discussion, a detailed analysis including the distribution of the hyperpolarized sample will be necessary which is beyond the scope of this work.

-


\section*{Summary and Conclusion}
Micron-scale NMR is at the cusp of reaching new milestones such as detecting NMR signals from microfluidics or from single cells.\cite{Glenn2018,Arunkumar2021,Smitseaaw7895,Bucher2020,Kehayias2017,bucher_quantum_2019} This work shows that for micron-scale NV-experiments the geometry of the diamond and the sample must be carefully chosen and, in the ideal case can even improve the signal size. We observe positive and negative contributions depending on the sample spin location within the sample volume on top of the diamond, leading to a strong dependence of the NMR signal size on the NV orientation with respect to the diamond surface (i.e. the diamond cut) and on the sample geometry. Whereas cone and spherical shapes increase, the cylindrical volume decreases the NMR signal, with important consequences for the detection of a single mammalian cell. In the case of NV-based hyperpolarization, the geometry of the sample nuclear spin distribution must be considered carefully.      \\

\section*{Acknowledgements}

This project was funded by the European Research Council (ERC) under the European Union’s
Horizon 2020 research and innovation program (grant agreement No 948049)” and by the Deutsche Forschungsgemeinschaft (DFG, German Research Foundation) - 412351169 within the Emmy Noether program.

\section*{Author contributions statement}
D.B.B conceived and supervised the study.  F.B. wrote the paper, made most simulations and analytic derivations. K.B. helped with analytic derivations and simulations. All authors discussed the results and contributed to the writing of the manuscript.

\section*{Additional information}

All authors declare that they have no competing interests.

\bibliography{main}
\bibliographystyle{plain}
\flushbottom

\section*{Supplementary material}
\subsection{Analytical derivation of the geometry factor}\label{sec:geometryfac}
Starting from Equation \ref{eq:signal}, we look at the part denoting the geometry factor, projected along the orientation of the NV center which is parallel to $\hat{B}_0 = (\sin{\gamma},0,\cos{\gamma})$:

\begin{equation}
\begin{split}
    G &=\int  \,  \left[ \frac{3 \hat{r} \left( \hat{r} \cdot \hat{m}_{\mathrm{max}} \right) -  \hat{m}_{\mathrm{max}}}{r^3}\right] \cdot \, \hat{B}_0 \, dV \\
    &= 3 \, \int  \frac{(\hat{r} \cdot \hat{B}_0) (\hat{r} \cdot \hat{m}_{\mathrm{max}})}{r^3} \, dV
\end{split}
\end{equation}
where we have used the orthogonality of $\hat{m}_{\mathrm{max}}$ and $ \hat{B}_0$. Integrating over a spherical cap in spherical coordinates with the center at the NV yields:

\begin{equation}
\begin{split}
    G &= 3 \, \int_{d_{NV}}^R \int_0^{\theta_{\mathrm{max}}} \int_0^{2\pi} \sin{\theta} \\ &\frac{(\sin{\gamma}\sin{\theta}\cos{\phi}+\cos{\gamma}\cos{\theta}) (\cos{\gamma}\sin{\theta}\cos{\phi}-\sin{\gamma}\cos{\theta})}{r} \, d\phi\,d\theta\,dr\\
    &= \pi \;\sin{\gamma}\;\cos{\gamma} \; \frac{2 R^3 -3 d_{NV} R^2 +d_{NV}^3}{R^3}
\end{split}
\end{equation}

where we used $\cos{\theta_{\mathrm{max}}}=d_{NV}/r$. Keeping $\theta_{\mathrm{max}}$ constant instead leads to an integration over a cone shaped volume. Using these boundary conditions, we end up with:

\begin{equation}
    G_{\mathrm{cone}}=\pi \;\sin{\gamma}\;\cos{\gamma}\;\sin{\theta_{\mathrm{max}}}^2\;\cos{\theta_{\mathrm{max}}} \; \log{\left(\frac{R}{d_{NV}}\right)}
\end{equation}

The analytical expressions for all other geometries can be derived by choosing the integration boundaries appropriately.

\subsection{Simulations of NV ensembles}\label{sec:NVlayers}
\begin{figure}[H]
\centering
\includegraphics[width=\linewidth]{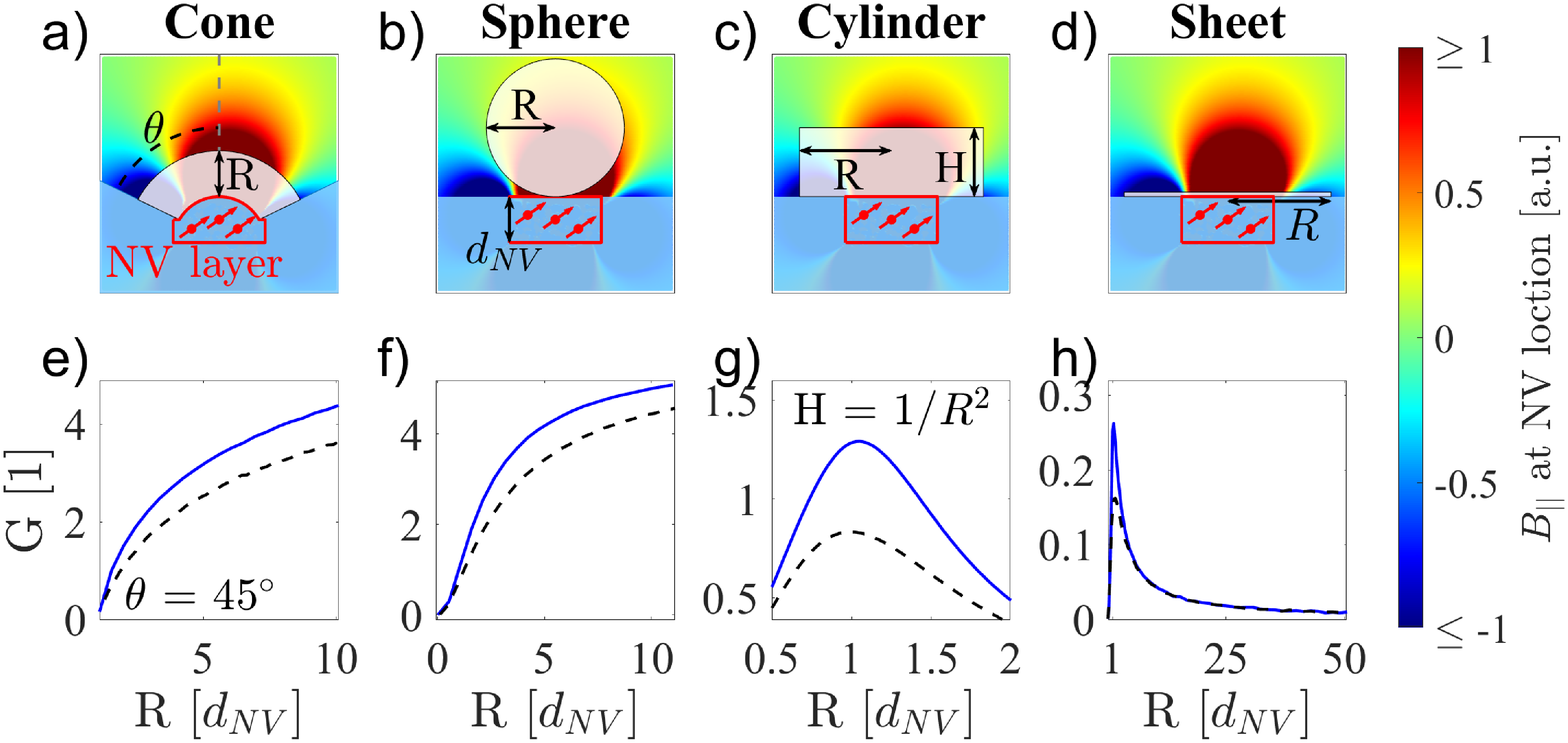}
\caption{\textbf{a)-d)} The setup for the four different geometries discussed in Figure \ref{fig:geometries}. Instead of one NV center at a distance $d_{NV}$ below the diamond surface, there are several NV-centers within a layer of thickness $d_{NV}$.\\
\textbf{e)-h)} The average geometry factor per NV center. The simulation data for the NV-ensemble is shown in blue and the data for a single NV at depth  $ d_{NV}$ is shown in black. The cone shaped, nearly-2D and spherical sample volume do not gain much from the ensemble. On the other hand, the geometry factor for the the cylindrical sample has a stronger dependence on $d_{NV}$, leading to a bigger change in the geometry factor compared to the single NV-center. }
\label{fig:geometries_layer}
\end{figure}

For detection of NMR signals, typically ensembles of NV-centers within a layer of defined thickness $d_{NV}$ are used. The numerical simulations shown above can easily be extended to layers containing multiple NV centers. For these simulations, the size of the NV-ensemble contributing to the signal was approximated as a cylinder of radius $R=d_{NV}$ and height $H=d_{NV}$, with NV-centers uniformly distributed in between, which is an reasonable estimate for typical $ d_{NV} \sim 10 \mu$m. Results for the geometries discussed in the main part of the paper can be found in Figure \ref{fig:geometries_layer}. 

\subsection{Monte Carlo simulation}\label{sec:MonteCarlo}
\begin{figure}[H]
\centering
\includegraphics[width=\linewidth]{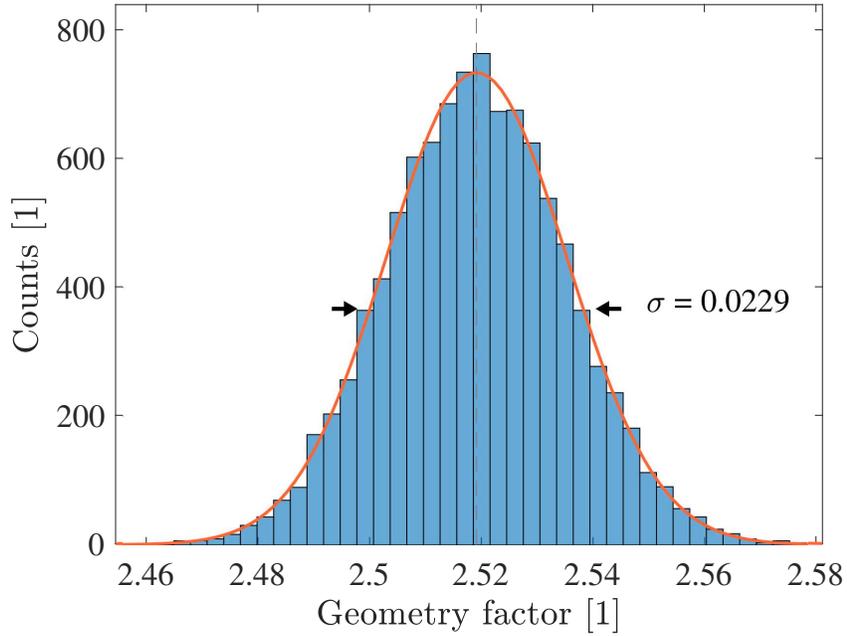}
\caption{
Results for $K=10^4$ individual simulations of a spherical cap of radius 10 $d_{NV}$ filled with N $\sim 3.6\cdot10^6$ sample-spins and one NV center at the origin. The orange curve is a gaussian fit to the simulated data with a standard deviation of $\sigma = 0.0091\;G$ with the center at G = 2.52 which is identical with the analytical value.
}
\label{fig:histogramm}
\end{figure}

Monte Carlo integration was performed to calculate the geometry dependence of a samples NMR signal at the position of the NV center. Simulations were repeated K times and averaged. For each simulation, sample dipoles were placed at $N_k$ random, uniformly distributed coordinates within the sample area, above the z = 1 plane. For simulations with one single NV center, the NV was positioned at the origin. For simulations of an NV layer $M_k$ random, uniformly distributed coordinates were created within an cylinder of radius 1 and height 1. The geometry factor was calculated by using the equation:

\begin{equation}
    G =1/K \; \sum_{k=1}^K \left[ \frac{V_{\mathrm{sample}}}{M_k\; N_k} \;\sum_{m=1}^{M_k} \left( \sum_{n=1}^{N_k}  \frac{3 \hat{r}_{m,n} \left( \hat{r}_{m,n} \cdot \hat{m}_{\mathrm{max}} \right) -  \hat{m}_{\mathrm{max}}}{r_{m,n}^3}  \cdot \hat{B}_0 \right) \right]
\end{equation}

where $r_{m,n}$ is the distance from sample-spin n to NV center m and $\hat{r}_{m,n}$ the normalized direction between the two. Typical values used in the simulations are N = $3.5\;10^6$, M = 1 and K = 100 for the single NV simulations and N = $1.5\;10^6$, M = 100 and K = 100 for the ensembles. Results for a Monte Carlo integration of a sample shaped like a spherical cap of radius 10 $d_{NV}$ can be found in Figure \ref{fig:histogramm}.

\end{document}